\documentclass[longbibliography,aps,prb,twocolumn,superscriptaddress,showpacs,amscd,amsmath,amssymb,verbatim]{revtex4-2}

\usepackage{graphicx}
\usepackage[OT1,T1]{fontenc}
\usepackage{revsymb}
\usepackage{physics}
\usepackage{bm}
\usepackage[dvipsnames]{xcolor}
\usepackage[separate-uncertainty=true,multi-part-units=single]{siunitx}
\usepackage[colorlinks=true,linkcolor=blue,citecolor=blue,urlcolor=blue]{hyperref}
\usepackage{soul}

\newcommand{\tbscnb}{Tb$_2$ScNbO$_7$}
\newcommand{\tbti}{Tb$_2$Ti$_2$O$_7$}
\newcommand{\tbhf}{Tb$_2$Hf$_2$O$_7$}
\newcommand{\tbsn}{Tb$_2$Sn$_2$O$_7$}
\newcommand{\przr}{Pr$_2$Zr$_2$O$_7$}
\newcommand{\prhf}{Pr$_2$Hf$_2$O$_7$}
\newcommand{\prsn}{Pr$_2$Sn$_2$O$_7$}
\newcommand{\prscnb}{Pr$_2$ScNbO$_7$}

\begin{document}

\title{Collective magnetic state induced by charge disorder in the non-Kramers rare-earth pyrochlore Tb$_2$ScNbO$_{7}$}

\author{Yann Alexanian}
\email[]{Yann.Alexanian@unige.ch}
\altaffiliation[]{current address: Department of Quantum Matter Physics, University of Geneva, 24 Quai Ernest-Ansermet, CH-1211, Geneva, Switzerland}
\affiliation{Institut N\'eel, CNRS and Universit\'e Grenoble Alpes, BP166, F-38042 Grenoble Cedex 9, France}
\affiliation{Institut Laue Langevin, CS 20156, 38042 Grenoble, France}

\author{Elsa Lhotel}
\affiliation{Institut N\'eel, CNRS and Universit\'e Grenoble Alpes, BP166, F-38042 Grenoble Cedex 9, France}

\author {Rafik Ballou}
\affiliation{Institut N\'eel, CNRS and Universit\'e Grenoble Alpes, BP166, F-38042 Grenoble Cedex 9, France}

\author{Claire V. Colin}
\affiliation{Institut N\'eel, CNRS and Universit\'e Grenoble Alpes, BP166, F-38042 Grenoble Cedex 9, France}

\author{Holger Klein}
\affiliation{Institut N\'eel, CNRS and Universit\'e Grenoble Alpes, BP166, F-38042 Grenoble Cedex 9, France}

\author{Antonin Le Priol}
\affiliation{Institut N\'eel, CNRS and Universit\'e Grenoble Alpes, BP166, F-38042 Grenoble Cedex 9, France}

\author{Flavien Museur}
\affiliation{Institut N\'eel, CNRS and Universit\'e Grenoble Alpes, BP166, F-38042 Grenoble Cedex 9, France}

\author{Julien Robert}
\affiliation{Institut N\'eel, CNRS and Universit\'e Grenoble Alpes, BP166, F-38042 Grenoble Cedex 9, France}

\author{Elise Pachoud}
\affiliation{Institut N\'eel, CNRS and Universit\'e Grenoble Alpes, BP166, F-38042 Grenoble Cedex 9, France}

\author{Pascal Lejay}
\affiliation{Institut N\'eel, CNRS and Universit\'e Grenoble Alpes, BP166, F-38042 Grenoble Cedex 9, France}

\author{Abdellali Hadj-Azzem}
\affiliation{Institut N\'eel, CNRS and Universit\'e Grenoble Alpes, BP166, F-38042 Grenoble Cedex 9, France}

\author{Bjorn F\aa{}k}
\affiliation{Institut Laue Langevin, CS 20156, 38042 Grenoble, France}

\author{Quentin Berrod}
\affiliation{Univ. Grenoble Alpes, CNRS, CEA, IRIG-SyMMES, 38000 Grenoble, France}

\author{Jean-Marc Zanotti}
\affiliation{Laboratoire L\'eon Brillouin (CEA-CNRS), Universit\'e Paris-Saclay, 91191 Gif-sur-Yvette, France}

\author{Emmanuelle Suard}
\affiliation{Institut Laue Langevin, CS 20156, 38042 Grenoble, France}

\author{Catherine Dejoie}
\affiliation{ESRF $-$ The European Synchrotron, CS40220, 38043 Grenoble, France}

\author {Sophie de Brion}
\affiliation{Institut N\'eel, CNRS and Universit\'e Grenoble Alpes, BP166, F-38042 Grenoble Cedex 9, France}

\author {Virginie Simonet}
\email[]{virginie.simonet@neel.cnrs.fr}
\affiliation{Institut N\'eel, CNRS and Universit\'e Grenoble Alpes, BP166, F-38042 Grenoble Cedex 9, France}
\date{\today}

\begin{abstract}
Geometrical frustration, as in pyrochlore lattices made of corner-sharing tetrahedra, precludes the onset of conventional magnetic ordering, enabling the stabilization of fluctuating spin states at low temperature. Disorder is a subtle ingredient that can modify the nature of these exotic non-ordered phases. Here, we study the interplay between disorder and magnetic frustration in the new pyrochlore \tbscnb\ where the non magnetic site presents a charge disorder Nb$^{5+}$/Sc$^{3+}$. Its quantification with sophisticated diffraction techniques (electrons, x-rays, neutrons) allows us to estimate the distribution of the splitting of the magnetic Tb$^{3+}$ non-Kramers ground state doublets and to compare it with excitations measured in inelastic neutron scattering. Combining macroscopic and neutron scattering measurements, we show that a spin glass transition occurs at 1~K while retaining strong spin liquid correlations. Our results suggest that \tbscnb\ stabilizes one of the novel disorder induced quantum spin liquid or topological glassy phases recently proposed theoretically.
\end{abstract}

\maketitle

\section{Introduction \label{sec:intro}}

Since a weak amount of structural disorder is generally an unavoidable ingredient in real materials, theoretical works have long attempted to take into account its influence in frustrated magnets where small perturbations can play a major role. This led to several fascinating predictions. In the Heisenberg antiferromagnet pyrochlore for instance, the role of disordered exchange interactions produced by random strains was shown to lead to a conventional spin glass transition at finite temperature. The transition temperature was proposed to scale with the amplitude of disorder or with magnetoelastic coupling strength \cite{Bellier2001,Saunders2007,Andreanov2010,Shinaoka2011}, while glassiness without disorder was also predicted \cite{Chandra1993,Cepas2012}. 

More recently, disorder has also been investigated within the most emblematic spin liquid phase stabilized on a pyrochlore lattice \cite{Gardner2010}, i.e. the spin ice state. It can be obtained in the case of Ising-like moments with local anisotropy axes along the $\langle111 \rangle$ directions of the tetrahedra combined with classical ferromagnetic interactions \cite{Harris1997}. This so-called Coulomb phase governed by local ice rules (two spins in and two spins out of each tetrahedron) has a massive ground state degeneracy with emergent magnetic monopole excitations \cite{Henley2010, Castelnovo2008}. Quantum fluctuations can melt this classical spin ice into a quantum spin liquid \cite{Onoda2010} where the degenerate ground states become entangled, and which carries deconfined and photon-like excitations \cite{Savary2012}. Sen and Moessner \cite{Sen2015} proposed that magnetic dilution produces a topological spin glass out of the spin ice state from the emergence of new interacting degrees of freedom, the impurity monopoles. In this spin glass, liquidity and glassiness compete leading to a loss of entropy at low temperature but retaining key features of the Coulomb phase such as the pinch points observed in neutron scattering \cite{Fennell2009}. In another spin ice model by Savary and Balents which addresses the case of non-Kramers magnetic ions \cite{Savary2017}, disorder takes the form of a random crystal field that breaks the $\mathcal{D}_{\mathrm{3d}}$ symmetry of the rare-earth site and thus splits the non-Kramers ground state doublet. The Hamiltonain is then equivalent to a random transverse field Ising model inducing quantum entanglement that leads to a quantum spin liquid, a Griffiths Coulomb glass or a spin glass, depending on the relative values of the interactions, the doublet splitting and its distribution \cite{Savary2017,Benton2018}. 
 
These predictions have been accompanied by experimental studies on various candidate materials among the pyrochlore family of formula $R_2B_{2}$O$_7$ where two interpenetrated pyrochlore lattices are occupied by rare-earth $R^{3+}$ ions on one hand and $B^{4+}$ ions on the other hand (Fig. \ref{fig:struct1}). A canonical spin glass has been reported in Y$_2$Mo$_2$O$_7$ where the magnetic ion is on the $B$ site \cite{Gingras1997,Gardner1999b,Booth2000}. Concerning pyrochlores with a magnetic rare-earth on the $R$ site, the presence of random strains has been quantified in Pr$_2$Zr$_2$O$_7$ and announced to be at the origin of its unconventional ground state with quantum spin liquid and quadrupolar correlation signatures \cite{Wen2017,Martin2017,Benton2018}. In Tb$_2$Ti$_2$O$_7$, the ground state is very sensitive to weak off-stoichiometry and magnetoelastic couplings \cite{Alexanian2023}, and it evolves from a spin liquid state to a quadrupolar order \cite{Taniguchi2013,Takatsu2016}. Moreover, substitution of the non-magnetic $B$ ion modifies the ground state toward a soft spin ice in Tb$_2$Sn$_2$O$_7$ \cite{Mirebeau2005,Dalmas2006} or a Coulomb phase coexisting with a spin glass state in Tb$_2$Hf$_2$O$_7$ \cite{Sibille2017,Anand2018}. 

To go further, it is crucial to carefully characterize the nature of the disorder, since it can affect the crystal field scheme of the rare-earth ions, the magnetic interactions, and also produce magnetic dilution, each of these ingredients having a specific influence. The study of materials with an intrinsic controlled disorder is therefore a promising way to unravel its role in the promotion of exotic quantum states. 

In this article, we focus on the charge disordered pyrochlore compound \tbscnb. The $R$ site is occupied by the non-Kramers ion Tb$^{3+}$ and the $B$ site is occupied by an {\it a priori} equal mixture of non-magnetic ions Nb$^{5+}$ and Sc$^{3+}$. The charge neutrality should impose a preferential equal occupation of +3 and +5 charges on each tetrahedron of the pyrochlore lattice, which is equivalent to a charge ice electronic material \cite{Anderson1956}. The present work investigates the influence of such controlled charge ice disorder on the magnetic properties of the Tb pyrochlore lattice.

\section{Disorder and structural properties of \tbscnb}

\begin{figure}[t!]
\centering
\includegraphics[width=\columnwidth]{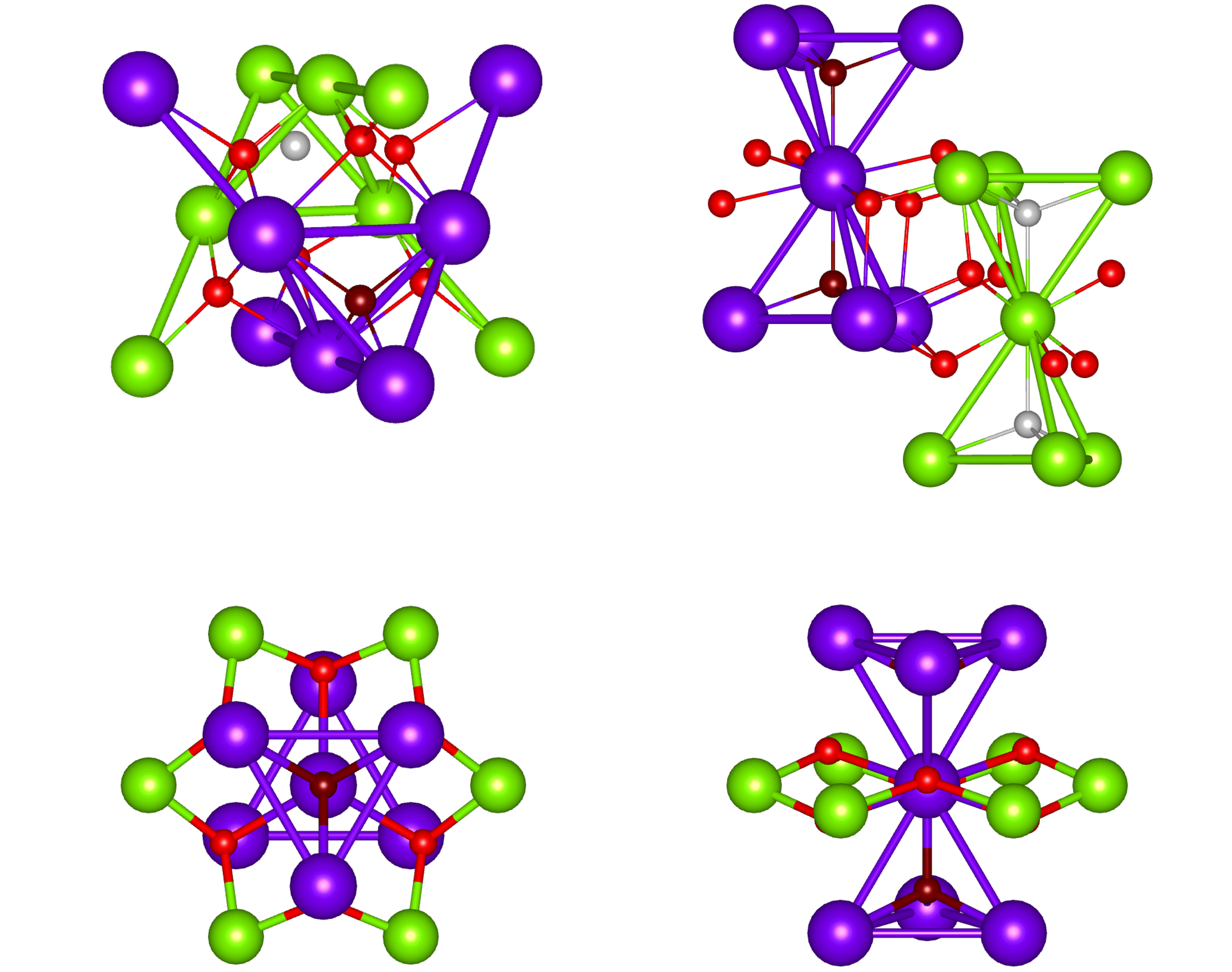}
\caption{Structure of \tbscnb: different views showing the two interpenetrating pyrochlore lattices made of Tb$^{3+}$ (purple) and Sc$^{3+}$ or Nb$^{5+}$ (green) ions respectively. The three possible oxygen sites 8b (dark red), 48f (red) and additional 8a (gray) are displayed, the latest one being empty in the perfect pyrochlore structure. The two bottom panels show that the environment of the Tb$^{3+}$ ions is constituted by an hexagon of six Sc$^{3+}$/Nb$^{5+}$ ions. }
\label{fig:struct1}
\end{figure}

\begin{figure}[t!]
\centering
\includegraphics[width=\columnwidth]{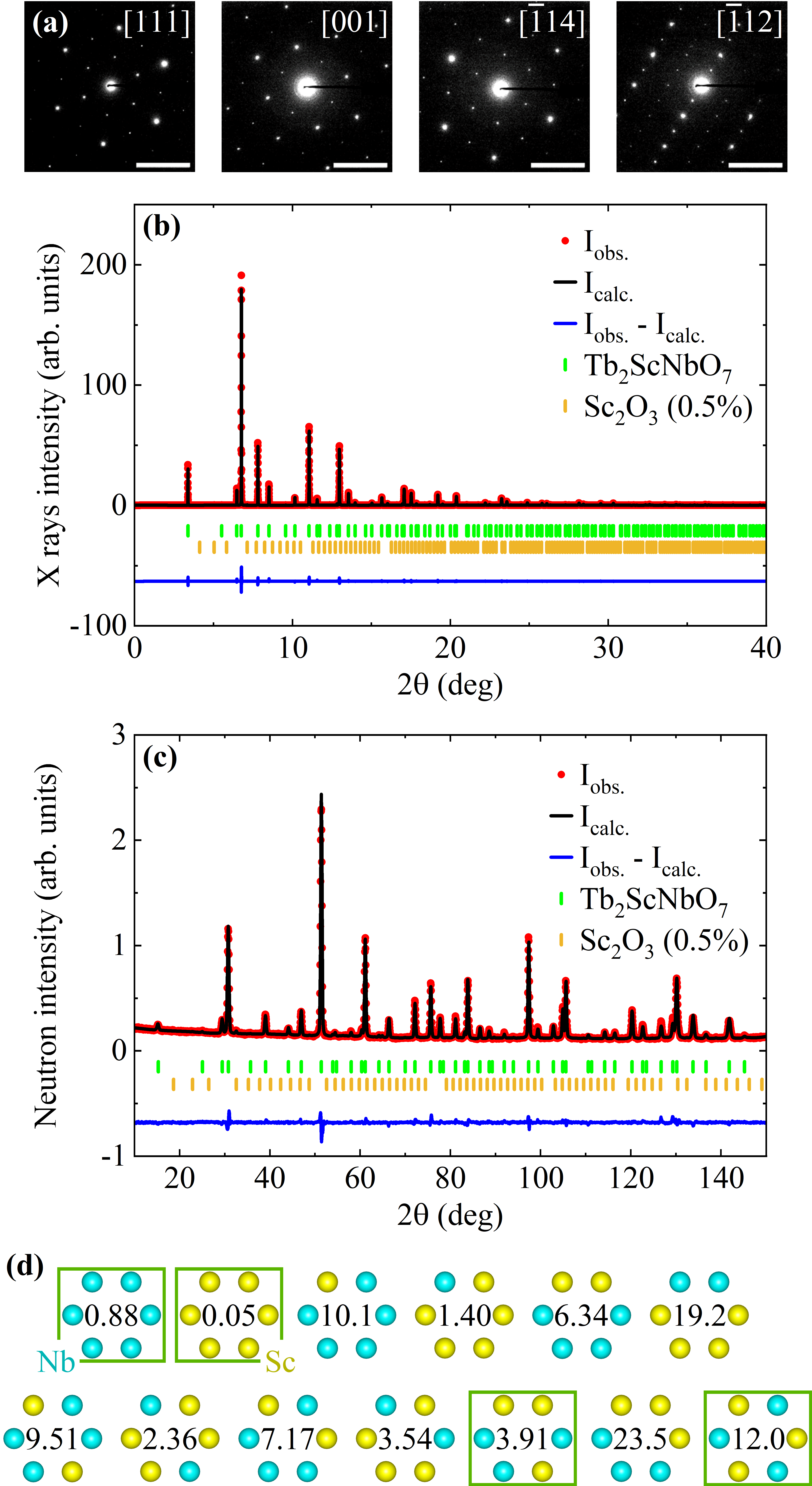}
\caption{Structural characterizations of \tbscnb: (a) Electronic diffraction patterns for different orientations of the micro-crystals measured with a transmission electron microscope at room temperature. The scale bar represents 5~nm$^{-1}$. X-ray (b) and neutron (c) diffractograms recorded on ID22 at ESRF and D2B at ILL respectively. The measurements on a polycrystalline sample at room temperature are in red, the joined Rietveld refinements are in black, the differences are in blue and the Bragg peak positions of the main phase and the impurity are indicated by the ticks. (d) Percentage of occurence of different environments of Tb$^{3+}$ with 57\% of Nb$^{5+}$ (blue) and 43\% of Sc$^{3+}$ (yellow), obtained from the statistical charge ice model \cite{Mauws2021}. Boxes indicate the non split Tb$^{3+}$ ground states.}
\label{fig:struct2}
\end{figure}

The \tbscnb\ powder sample has been prepared by solid state reaction (Appendix \ref{AppendixA}). Laboratory x-ray diffractograms have ensured the high quality of the sample with few non-magnetic impurities. The structural refinement has confirmed the cubic pyrochlore structure with space group Fd$\bar{3}$m, lattice parameter $a$=10.3935(5)~\AA , and the 48f oxygen coordinate $x=0.3370(2)$ (Appendix \ref{AppendixB}). Additional electronic diffraction has been performed on tiny crystallites of the same batch. The diffraction patterns in Fig. \ref{fig:struct2}a show Bragg peaks, all indexed in the pyrochlore structure, with no substructure additional peaks that would have indicated a NbSc long-range ordering tendency. 

The main defects reported in pyrochlore compounds are the cationic antisite defects (exchange of $R^{3+}$ and $B^{4+}$ cations) and Frenkel defects \cite{Minervini2000}. The latter are oxygen vacancies on 48f sites associated with additional oxygens on 8a sites (gray spheres in Fig. \ref{fig:struct1}). These defects become more numerous when the ratio of the $R^{3+}$ over $B^{4+}$ ionic radius is close to 1.46, the value below which the pyrochlore structure is destabilized towards the fluorite one. This ratio is equal to 1.50 for the average Nb$^{5+}$/Sc$^{3+}$ ionic radius in the stability region of the pyrochlore structure \cite{Gardner2010,Ortiz2022}. 

\begin{figure*}
\centering
\includegraphics[width=\textwidth]{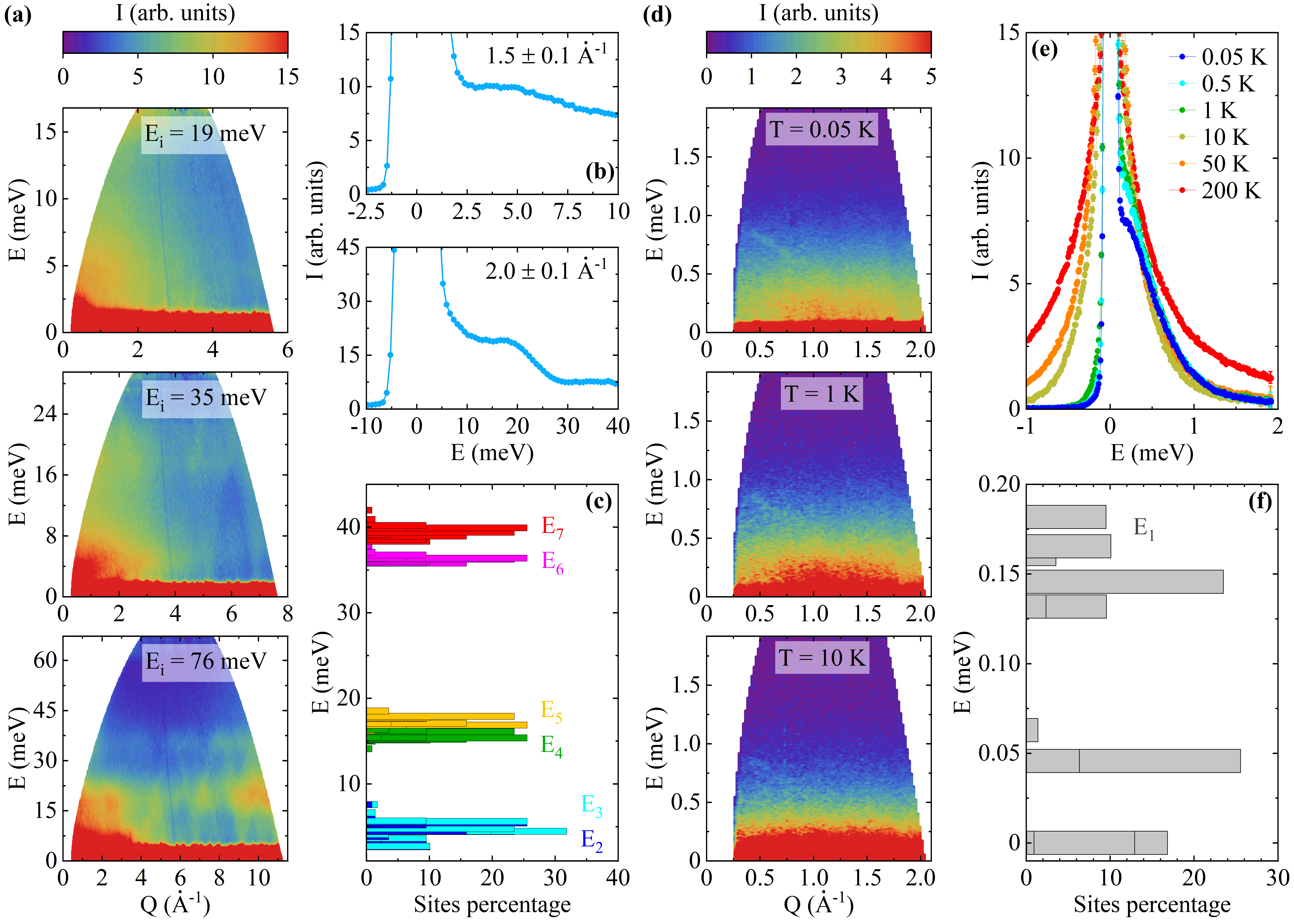}
\caption{(a) Energy $E$ versus scattering vector $Q$ maps of scattered neutron intensity recorded on PANTHER for 19, 35 and 76~meV incident energies at 1.5~K. (b) Constant energy cuts performed at $Q=1.5\pm0.1$ and $2\pm0.1$~\AA$^{-1}$, for $E_{\mathrm{i}}=19$ and 76~meV respectively. (c) Energy distribution of the CEF levels E$_{2-7}$ calculated from a point charge model (see text). (d) Energy versus scattering vector maps of neutron intensity recorded on SHARP at different temperatures with an incident wavelength of 5.1~\AA. A measurement using a 4.1~\AA\ wavelength allowed us to check the absence of signal around 1.5~meV. (e) $Q$-integrated neutron intensity between 0.55 and 1.65~\AA$^{-1}$ versus energy for different temperatures. (f) Calculated histogram of the ground doublet energy splitting (E$_1$ level) for the different Tb$^{3+}$ environments in the presence of 57\% Nb / 43\% Sc disorder. Contributions to the same splitting from different environments are cumulated inside a single bar and separated by black lines.}
\label{fig:PANTHER}
\end{figure*}

To characterize the possible defects and disorder in \tbscnb, we combined high-resolution synchrotron x-ray and neutron diffraction measurements (Appendix \ref{AppendixA}), shown in  Fig.~\ref{fig:struct2}b and \ref{fig:struct2}c, respectively. The two diffractograms were simultaneously refined with the Fullprof suite {\cite{FullProf1993} taking advantage of the complementarity of the techniques (sensitivity of neutrons to oxygens and X-ray scattering contrast between Tb and Nb). The refinement shows no evidence for cationic antisite disorder. Meanwhile, an excess of Nb$^{5+}$ (57\%) compared to Sc$^{3+}$ (43\%) is observed, leading to different charge environments of the Tb$^{3+}$ ions (Fig. \ref{fig:struct2}d). This positive charge excess is compensated by additional negatively charged oxygens on the 8a sites (up to 17\% occupancy, see table \ref{TableStruct} in Appendix \ref{AppendixB}) without any depletion of the 48f sites, ruling out a significant presence of Frenkel defects. 

\section{Tb$^{3+}$ single-ion Crystal Electric Field in \tbscnb}

Structural disorder is expected to play a role in the single-ion magnetic properties, in particular through its influence on the crystal electric field (CEF) scheme of non-Kramers ions such as Tb$^{3+}$. In \tbscnb\ with local $\mathcal{D}_{\mathrm{3d}}$ symmetry, the CEF splits the ground state manifold into 4 doublets and five singlets. The $B$ ion charge disorder might modify the energy of all levels and lift the doublets' degeneracy \cite{Rau2019}.

We explored the \tbscnb\ CEF scheme by inelastic neutron scattering (Appendix \ref{AppendixA}). The excitation spectrum at 1.5~K (Fig. \ref{fig:PANTHER}a) shows strongly dispersive modes visible up to 35~meV that can be ascribed to phonons. In addition, CEF levels are identified as non-dispersive signals broadened by charge disorder. Their magnetic nature is assessed by their $Q$ dependence, which follows the squared Tb$^{3+}$ magnetic form factor at low $Q$ (Appendix \ref{AppendixC}). The two first excited levels are located around 4.5 and 17 meV (Fig. \ref{fig:PANTHER}b). Another CEF level might be present around 35~meV although largely mixed with phonons. 

We attempted to reproduce the experimental CEF scheme and its broad distribution using a point charge model \cite{Sternheimer1968} assuming 57\% of Nb$^{5+}$ and 43\% of Sc$^{3+}$ on the $B$ site and a charge ice rule. The coordination polyhedron of Tb$^{3+}$ consists in eight oxygens (48f and 8b), an hexagon of further-neighbor $B$ site ions and six Tb$^{3+}$ (Fig.~\ref{fig:struct2}c). This results in 13 Nb$^{5+}$/Sc$^{3+}$ arrangements on the hexagon, already reported for Nd$_2$ScNbO$_7$ \cite{Mauws2021,Scheie2021}, and shown in Fig.~\ref{fig:struct2}d with their percentage of occurence. 
The energy spectrum for the 13 configurations was computed (Appendix \ref{AppendixD}), which gives the CEF energy scheme displayed in Fig~\ref{fig:PANTHER}c. The first two levels are distributed around 4.5 and 16.2~meV while the next ones are around 38~meV. This is satisfactory regarding the crudeness of the point charge model, the few accessible observables and the broadness of the crystal field levels. 

The charge disorder also has a strong impact on the ground CEF level, since 83\% of the charge configurations produces a doublet splitting. The calculated splitting distribution spreads between 0 and 0.2~meV, with more levels around 0.15~meV (Fig. \ref{fig:PANTHER}f). These results can be compared with low-energy inelastic neutron scattering measurements (Fig. \ref{fig:PANTHER}d,e). At high temperature, a quasielastic signal whose width decreases when decreasing the temperature is ascribed to a usual slowing down of the single-ion spin dynamics, due to decreasing interactions with phonons. It transforms below 10~K into a signal at low energy that appears as a shoulder around 0.15~meV with a broad tail in energy cuts. This could not be reproduced by any combination of a quasielastic signal and a single inelastic excitation with a finite lifetime (Appendix \ref{AppendixC}), and rather seems to result from multiple inelastic peaks, reflecting the CEF splitting distribution discussed above. However, the measured excitation extends well beyond the largest calculated energy gap of 0.2~meV, and in addition, the low energy signal is slightly modulated in $Q$ with a maximum of intensity around 1.2~\AA$^{-1}$. This reflects the limitations of our model: (i) it only considers the $B$ site ion charge distribution but not the resulting structural distortions, nor the charge disorder due to additional 8a oxygens; (ii) it is a single ion picture, which does not account for the cooperative behavior associated with the magnetic interactions, as discussed below. 

\section{Beyond single ion in macroscopic measurements}

\begin{figure}
\centering
\includegraphics[width=\columnwidth]{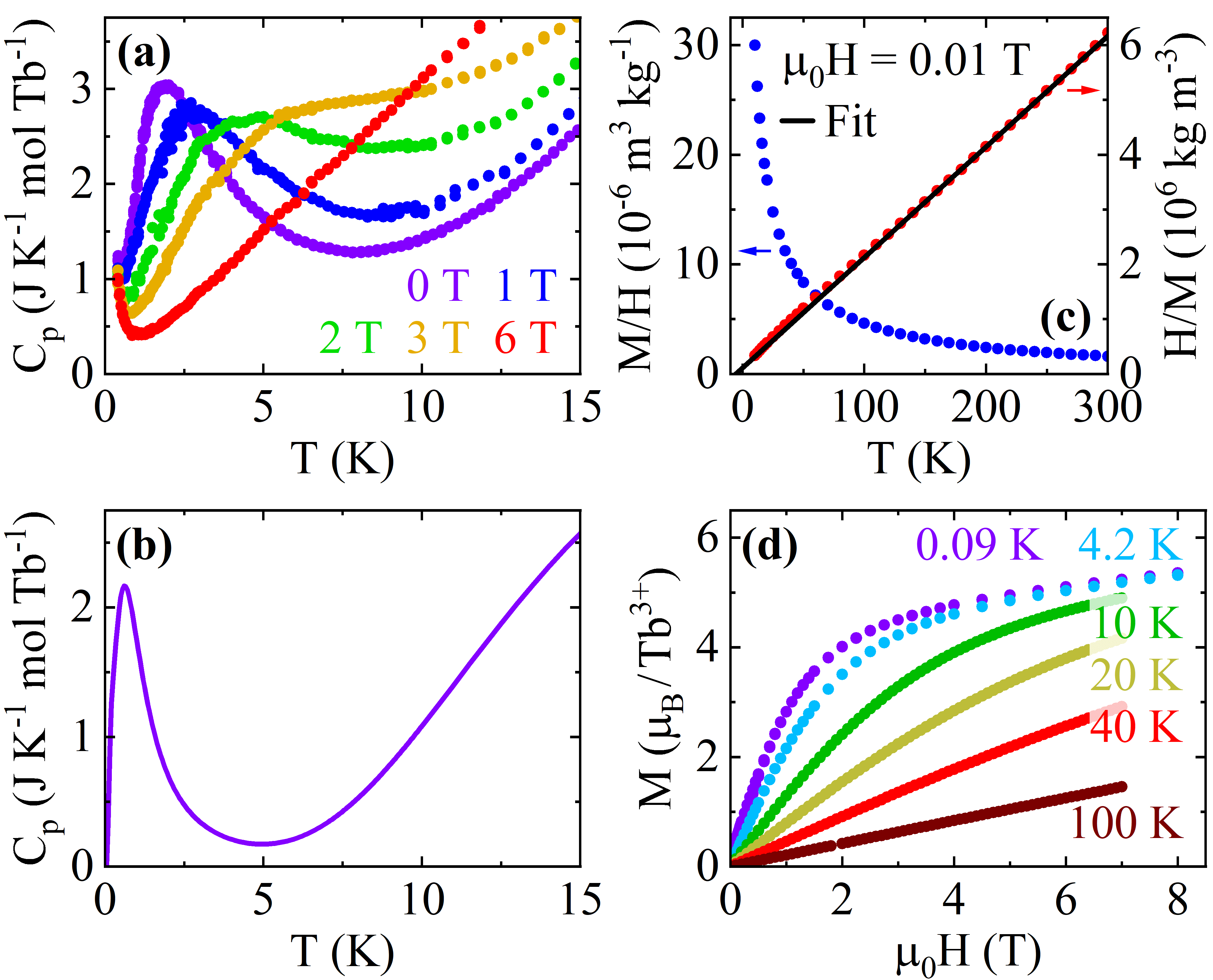}
\caption{(a) Specific heat versus temperature measured for different magnetic fields. (b) Calculated specific heat using the point charge model with charge disorder. (c) Magnetic susceptibility (blue dots) and its inverse (black dots) versus temperature measured in 0.01~T. The red line is a Curie-Weiss fit $1/\chi=(T-\theta)/C$ between 150 and 300~K with the Curie-Weiss temperature $\theta$ and the Curie constant $C \propto \mu_{\rm eff}^2$ where $\mu_{\rm eff}$ is the effective moment (black line). (d) Magnetization versus magnetic field for different temperatures.}
\label{fig:Cpchi}
\end{figure}

The experimental details on all macroscopic measurements are given in Appendix \ref{AppendixA}.
Specific heat measurements show no ordering transition down to 400 mK (Fig. \ref{fig:Cpchi}a). An upturn below 1~K is attributed to a nuclear Schottky anomaly resulting from the splitting of the energy levels of the $^{159}$Tb nuclear spins by the hyperfine field of the Tb$^{3+}$ electronic moment. The most interesting feature is a broad maximum observed around 2~K, which could arise from a ground-state doublet splitting. Our calculations in the disordered point charge model (Fig. \ref{fig:Cpchi}b) indeed produce the expected peak, which is however narrower and located at lower temperature than experiment pointing out again that additional ingredients (structural disorder, magnetic correlations \cite{Gingras2000}) are necessary to describe quantitatively the \tbscnb\ low temperature properties.

The \tbscnb\ magnetic state was further explored through magnetization measurements (Fig. \ref{fig:Cpchi}c). A fit of the susceptibility to a Curie-Weiss law between 150 and 300~K yields an effective moment of 9.5 $\mu_{\rm B}$ close to the expected value for free Tb$^{3+}$ (9.72 $\mu_{\rm B}$) and a Curie-Weiss temperature, $\theta=-5.5$~K, slightly smaller than in other Tb pyrochlores \cite{Gardner1999a,Matsuhira2002,Sibille2017,Anand2018}. However, CEF effects are known to affect the susceptibility \cite{Gingras2000}, as confirmed by our single-ion calculations of the inverse susceptibility for the 13 charge disorder configurations (Appendix \ref{AppendixD}), so that the interactions strength cannot be deduced from the Curie-Weiss fit. Several authors attempted to determine the intrinsic contribution of the magnetic interactions (exchange and dipolar) to the Curie-Weiss temperature in Tb$_2$Ti$_2$O$_7$ and Tb$_2$Sn$_2$O$_7$ and they found values ranging between -7 and -20 K \cite{Gingras2000,Mirebeau2007,Takatsu2016}.

Magnetization curves versus magnetic field are shown in Fig. \ref{fig:Cpchi}d. At 90~mK, the magnetization does not saturate up to 8~T, and it reaches 5.3 $\mu_{\rm B}$ per Tb$^{3+}$ ion, lower than the free ion value $g_JJ\mu_{\rm B}=9$ $\mu_{\rm B}$, but close to the values reported for other Tb pyrochlores \cite{Yasui2002, Mirebeau2007, Legl2012, Lhotel2012, Matsuhira2002, Singh2022, Anand2018}. This is in agreement with the calculated ground state doublet wavefunctions for the 13 charge distributions, prominently comprised of $\ket{J^z=\pm 5}$. 

ZFC-FC magnetization measurements were performed down to 80~mK in various magnetic fields (Fig. \ref{fig:ZFCchiAC}a). The two curves display a bifurcation below 1~K at the lowest field. This behavior is the signature of spin freezing, which is supported by the opening of an hysteresis cycle in the magnetization curves below 1~K (Appendix \ref{AppendixE}). The freezing is only partial: an estimation of the frozen spins, given by $1-\chi_{\rm{ZFC}}/\chi_{\rm{FC}}$, levels at $\approx$ 50 \%. This fraction concerns the spins whose dynamics is slower than the time scale of the measurement after applying a field, typically 1 min. The freezing temperature and the fraction of frozen spins decrease when increasing the magnetic field. The spins can indeed more easily reach their equilibrium state when the Zeeman energy associated with the applied field increases.

\begin{figure}
\centering
\includegraphics[width=\columnwidth]{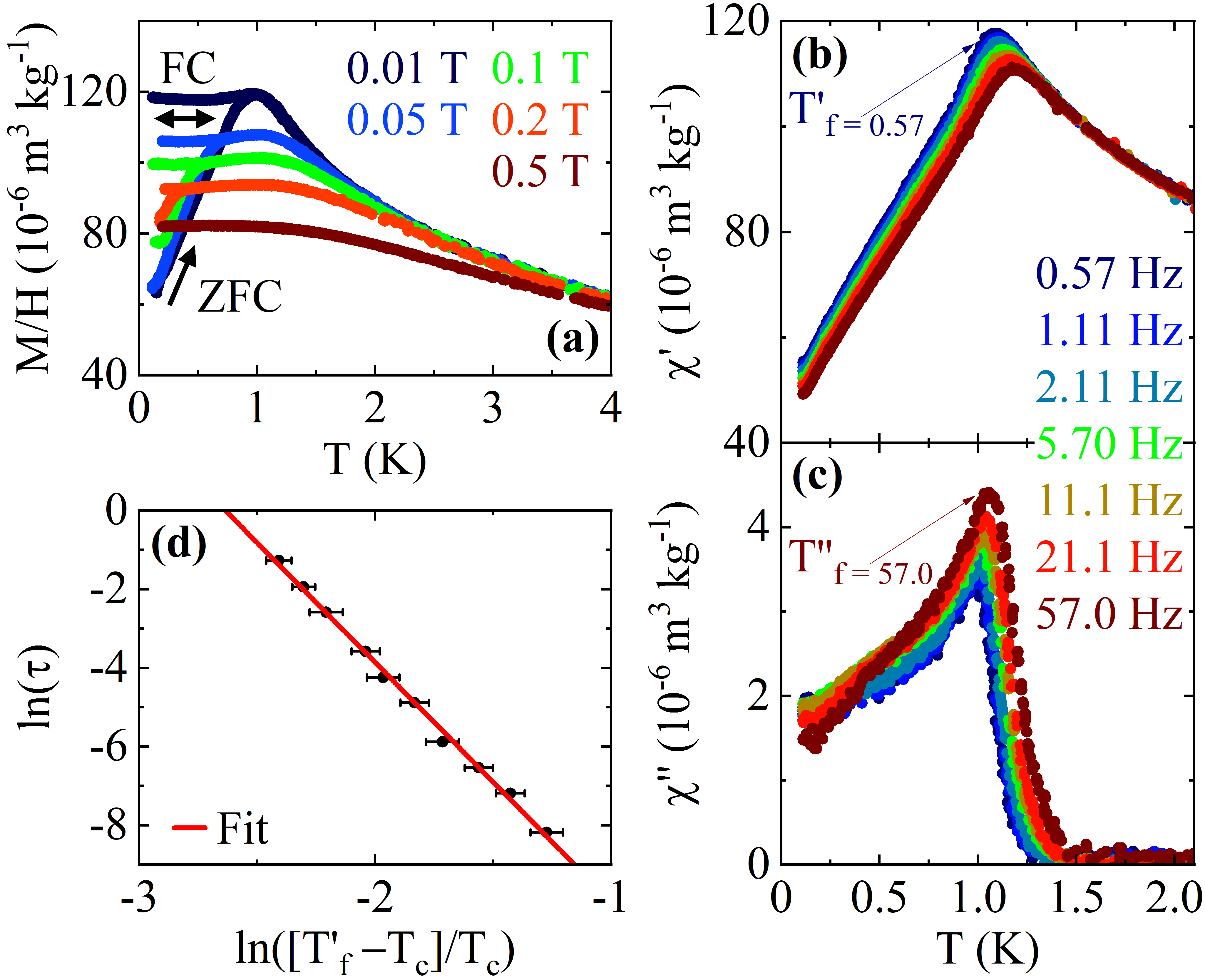}
\caption{(a) Low temperature Zero Field Cooled (ZFC) and Field Cooled (FC) magnetization for different magnetic fields. (b-c) AC susceptibility: real $\chi'$ (b) and imaginary $\chi''$ (c) parts as a function of temperature for different frequencies; (d) $\ln \tau=1/2\pi f$ versus $\ln\left([T'_f-T_c]/T_c\right)$ (black dots), with $T_c=1$~K. The red line is a fit to the spin glass dynamical scaling law (see text). $T'_f$ ($T''_{f}$) is the temperature of the maximum of $\chi'$  ($\chi''$). }
\label{fig:ZFCchiAC}
\end{figure}

We further investigated the freezing behavior through AC susceptibility measurements. The real part of the susceptibility $\chi'$ exhibits a peak around 1~K whose position $T'_{f}$ shifts to higher temperature for higher frequencies and is associated with a dissipative peak in the imaginary part $\chi''$ (Fig. \ref{fig:ZFCchiAC}b,c). An Arrhenius law yields unphysical values of the relaxation times ruling out a thermally activated process to explain the observed irreversibilities (Appendix \ref{AppendixE}). On the contrary, the Mydosh parameter $\phi=\Delta T'_f/(T'_f\Delta \log f)\approx 0.05$, which describes the peak shift with frequency, has a typical value for insulating spin glasses \cite{Mydosh1993}. The $T'_f$ dependence of the relaxation time $\tau$ (Fig. \ref{fig:ZFCchiAC}d) is indeed accounted for in the frame of a spin glass transition described by the dynamical scaling law: $$\tau=\tau_0\left({T'_f-T_c\over T_c}\right)^{-z\nu}$$ where $T_c$ is the spin-glass transition temperature, while $z$ and $\nu$ are the dynamical and correlation length critical exponents respectively \cite{Souletie1985}. The parameter $z\nu=6.1\pm0.1$ is in the expected range for a canonical spin-glass transition \cite{Mydosh1993,Lago2012,Banerjee2023,Malinowski2011}. $\tau_0=1.1\pm0.1\times10^{-7}$ s is consistent with relaxation times reported in the Tb$_2$Hf$_2$O$_7$ pyrochlore \cite{Sibille2017,Anand2018} but is lower than usual values for spin glasses, suggesting that the interacting entities are not spins but rather spin clusters in our case \cite{Mukadam2005,Malinowski2011}. Finally, $T_c=1$~K coincides with the freezing temperature obtained from magnetization measurements and the spin glass picture is also consistent with the observed decrease of the freezing temperature while increasing the magnetic field (Fig.~\ref{fig:ZFCchiAC}a). 

\begin{figure*}
\centering
\includegraphics[width=\textwidth]{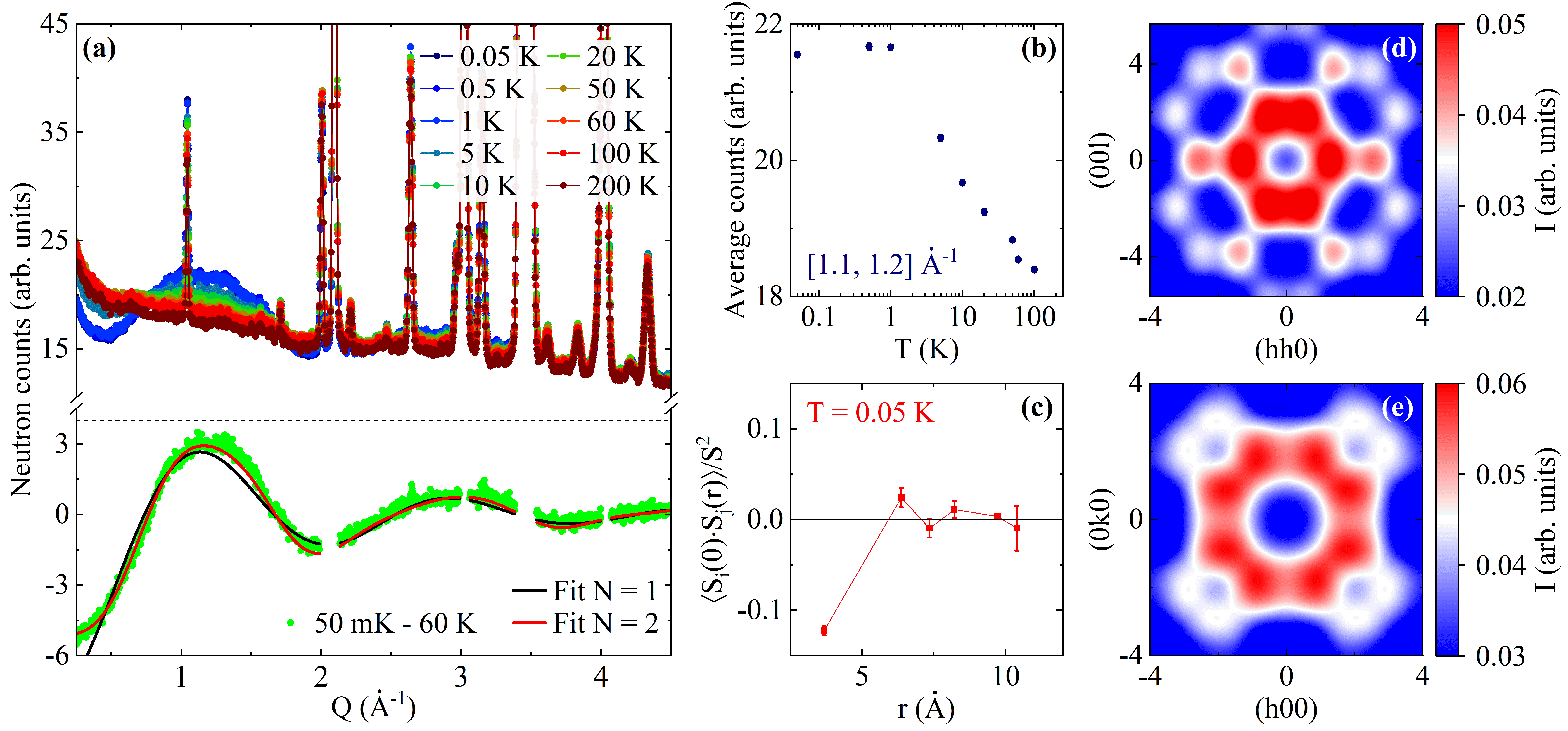}
\caption{Neutron diffraction of \tbscnb\ recorded on D1B: (a) (top) Temperature dependence of the powder neutron diffractograms between 200~K and 50~mK showing the rise of the magnetic diffuse scattering and (bottom) difference of the 50~mK and 60~K diffractograms isolating the magnetic diffuse scattering (green). The black and red lines show the fits obtained with {\sc spinvert} considering $N=1$ and 2 (see text) respectively. (b) Magnetic diffuse scattering at the first maximum integrated between 1.1 and 1.2~\AA$^{-1}$ as a function of temperature; (c) Normalized magnetic correlation function $\left<\bm{S}_{i}(0)\cdot\bm{S}_{j}(\bm{r})\right>/S^{2}$ for the successive Tb$^{3+}$ neighbors as a function of the distance, extracted with the program \textsc{spincorrel} \cite{Paddison2013} for $N=2$. Note that the first and second neighbor distances are 3.68 and 6.37 \AA\ respectively. (d,e) Neutron diffuse scattering in the (d) $(hhl)$ and (e) $(hk0)$ scattering planes reconstructed from RMC calculations.}
\label{fig:D1B}
\end{figure*}

\section{Neutron diffuse scattering}

The macroscopic studies on \tbscnb\ presented above point out a disordered correlated state which undergoes a partial spin glass transition at $T_c=1$ K. Neutron diffraction is a key tool to probe the associated magnetic correlations. In our experiment (Appendix \ref{AppendixA}), no magnetic Bragg peaks are observed down to 50~mK confirming the absence of long-range order (Fig. \ref{fig:D1B}a). Instead, a diffuse magnetic scattering, with broad peaks at 1.2 and 3.2~\AA$^{-1}$ characteristic of short-range magnetic correlations, starts to rise below 60~K. It becomes temperature independent below $T_c$, as shown by the temperature dependence of the first broad peak intensity (Fig. \ref{fig:D1B}b). This diffuse scattering and its temperature evolution are very similar to what was reported in Tb$_2$Ti$_2$O$_7$ \cite{Gardner1999a, Gardner2001} and Tb$_2$Hf$_2$O$_7$ \cite{Sibille2017, Anand2018}, where it was interpreted as the signature of spin liquid correlations due to magnetic interactions.

To get insights into these correlations, we have performed reverse Monte-Carlo simulations using the program \textsc{spinvert} \cite{Paddison2013} with $N \times N \times N$ unit cells, each containing 16 isotropic spins. A very good agreement between the measured and calculated diffuse scattering is already achieved for $N=2$ (Fig.~\ref{fig:D1B}a), which is not improved by increasing $N$ further. The obtained correlations are antiferromagnetic for first neighbors and decrease very fast, oscillating around zero for further neighbors (Fig. \ref{fig:D1B}c). The reconstructed diffraction pattern in the $(hhl)$ scattering plane (Fig.~\ref{fig:D1B}d) is qualitatively consistent with the one measured on a \tbhf\ single crystal \cite{Sibille2017}. Some fine details such as the pinch-points are not reproduced due to powder average and small system size. We nevertheless anticipate that the same kind of spin liquid and Coulomb phase signatures are present in \tbscnb, which could be investigated on a single crystal. 

\section{Discussion and conclusion}

We showed that the new \tbscnb\ pyrochlore oxide gathers striking features emerging in frustrated system models with disorder that predict original glassy states: signature of a well-defined canonical spin glass transition around 1~K, low energy collective excitations and spin-liquid like correlations. This results from 57\%~Nb$^{5+}$/43\%~Sc$^{3+}$ charge disorder on the $B$ site, whose main consequence is the splitting of the Tb$^{3+}$ non-Kramers ground state doublet. Because of time-reversal symmetry, this acts as a random transverse field \cite{Savary2017} providing quantum fluctuations. In \tbscnb, the distribution of splitting, proportional to the random field, is reflected in the low energy signal measured by inelastic neutron scattering (Fig.~\ref{fig:PANTHER}e). Its width is at least twice as large as than the average splitting roughly estimated to $\approx 0.15$~meV, itself smaller than magnetic interactions as estimated in other Tb pyrochlores \cite{Takatsu2016,Gingras2000,Mirebeau2007}. \tbscnb\ could then stabilize an exotic quantum frozen state arising from the confinement of the $U(1)$ gauge theory (see phase diagram of Fig.~1 in Ref. \cite{Benton2018}). 

The conditions to obtain the above features appear, however to be rather drastic. In Tb$_2$InSbO$_7$, where the $B$ site is also occupied by a mixture of different charged ions In$^{3+}$ and Sb$^{5+}$ \cite{Ortiz2022}, the disorder is too strong, due to a large difference in their ionic radii ($r_{{\mathrm{In}}^{3+}}$=0.8 \AA, $r_{{\mathrm{Sb}}^{5+}}$=0.6 \AA\ compared to $r_{{\mathrm{Sc}}^{3+}}$=0.745 \AA, $r_{{\mathrm{Nb}}^{5+}}$=0.64 \AA). No clear spin glass transition is observed, similarly to other Tb pyrochlores with some level of disorder \cite{Gaulin2015}, with the exception of \tbhf\ \cite{Sibille2017}. This last system presents strong similarities with \tbscnb\ including a spin glass transition around 700~mK and a diffuse scattering characteristic of a Coulomb phase. The ground state splitting is roughly estimated to 0.04-0.4 meV, in the same range as for \tbscnb. In addition, it has a first excited CEF level around 4~meV as in \tbscnb\ and contrary to \tbti\ and \tbsn\ where a low energy CEF level at 1.5~meV is observed \cite{Mirebeau2007}. This large separation between the ground state doublet and first excited state, allowing the low temperature properties to be dictated solely by the ground state CEF doublet, appears then to be necessary in the realization of these exotic disordered phases. Interestingly, the exact nature of the disorder does not seem to be so crucial  - provided it is not too strong and that it induces a ground state splitting. It is indeed rather different in \tbscnb\ and \tbhf. In the former, the additional 8a oxygens compensating the charge imbalance are out of the magnetic superexchange paths between the Tb$^{3+}$ ions and they should induce less exchange randomness than the Frenkel pair defects detected in Tb$_2$Hf$_2$O$_7$. 

It is useful to compare these results with Pr based pyrochlores since Pr$^{3+}$ is also a non-Kramers ion. In \przr, random strains were shown to induce a wide ground state splitting distribution maximum at about 0.4~meV \cite{Martin2017,Wen2017}, which competes with the magnetic interactions, resulting in a complex spin liquid state with quadrupolar correlations. The study of the charge disorder compound Pr$_2$NbScO$_7$ thus appears of great interest (Appendix \ref{AppendixF}). However, our macroscopic measurements suggest that this system is in a trivial paramagnetic state of diluted magnetism, which is confirmed by the absence of a correlated magnetic neutron diffuse scattering down to 50~mK. This can be explained by considering the same point charge model as used for \tbscnb. The obtained ground-state splitting is much larger (3-4~meV in average) than in the former compound, and overall significantly larger than the magnetic interactions. The latter are thus too weak to mix the split doublet wavefunctions of 83\% of Pr$^{3+}$ ions, which thus remain in a non magnetic singlet state. 

In conclusion, whereas the Nb$^{5+}$/Sc$^{3+}$ disorder on the non-magnetic site produces a non cooperative trivial state in \prscnb, it creates an exotic strongly correlated ground state in \tbscnb, which might be one promising realization of frustration driven unconventional frozen spin liquid states predicted by recent theories \cite{Savary2017, Benton2018, Sen2015}. Our study demonstrates a way towards the realization of these disorder-induced complex phases with competing liquid-glass behaviors, and how this requires a delicate balance between the distribution of the ground state doublet splitting and the strength of the magnetic interactions. The observed excess of Nb$^{5+}$ ions over the Sc$^{3+}$ ones in \tbscnb\ suggests, by varying this ratio, a way to tune the splitting distribution and further control the magnetic ground state.

\acknowledgements
We would like to thank Sofjen Djelit for his technical support for the D1B experiment, Jo\"el Balay for his instrumental support for the synthesis of the samples and Pierre Lachkar for his help during the specific heat measurements. We thank C. Paulsen for allowing us to use his SQUID dilution magnetometers. We acknowledge financial support from the “Agence Nationale de la Recherche” under Grant No. ANR-15-CE30-0004.

\appendix
\section{Experimental details}
\label{AppendixA}

Polycrystalline samples of $R_2$ScNbO$_7$ ($R=$Tb, Pr) were synthesized by solid-state reaction as in Ref. \cite{Zouari2008}. Cationic stoichiometric amounts of the precursor oxides (Tb$_4$O$_7$, Pr$_6$O$_{11}$, Sc$_2$O$_3$, Nb$_2$O$_5$) were ground and heated in air for several days at 1500$^{\circ}$C with intermediate re-grindings. The structure and purity of the samples were checked by powder X-ray diffraction using a Philips diffractometer with Cu $K_{\alpha}$ wavelength and by electron diffraction in a Philips CM300 transmission electron microscope operated at 300~kV and equipped with a TVIPS F416 CMOS camera at the Institut N\'eel. Less than 1\% of PrNbO$_4$ and Sc$_2$O$_3$ was detected in \prscnb\ and \tbscnb\ respectively. 
In addition, for \tbscnb, neutron diffraction was measured on the high resolution powder diffractometer D2B at the high flux reactor of the ILL with a wavelength of 1.594 ~\AA\ (\href{https://dx.doi.org/10.5291/ILL-DATA.EASY-792}{doi:10.5291/ILL-DATA.EASY-792}), and high-resolution X-ray diffraction was measured on the ID22 beamline at the European Synchrotron Radiation Facility (ESRF) using a wavelength of 0.354172(4)~\AA. 

Specific heat measurements were performed on a Quantum Design Physical Property Measurement System (PPMS) with a $^3$He insert between 0.4 and 20~K. Samples were pressed heated pellets stuck to the PPMS puck with apiezon grease, with a mass of 1.58 and 3.43~mg for the Tb and Pr samples, respectively. The addenda contribution was measured apart and removed from the total specific heat measured with the sample. 

Magnetometry measurements were performed by the extraction method, using two SQUID magnetometers, a MPMS Quantum Design one in the 2-300~K range and another one equipped with a miniature dilution refrigerator developed at the Institut N\'eel for measurements between 90~mK and 4.2~K \cite{Paulsen2001}. In addition, AC susceptibility measurements were performed in the latter, with frequencies of the AC magnetic field between 0.057 to 570~Hz and magnitudes between 0.09 and 0.18~mT. Powder samples of mass 145 mg and 20.6 mg, wrapped in plastic films and inserted in a food straw, were used for the susceptibility and magnetization measurements in the MPMS magnetometer.  For the low-temperature measurements, powder samples of mass 8.6 and  7.2 mg for the Tb and Pr samples respectively, were mixed with apiezon grease and inserted in a copper pouch to ensure a good thermalization. 

Further powder neutron scattering measurements were performed on the \tbscnb\ sample ($m\approx 2$ g). Neutron diffraction was measured on D1B (Néel CRG@ILL) using a wavelength of 2.52 \AA\ and using a dilution fridge (\href{https://dx.doi.org/10.5291/ILL-DATA.5-31-2717}{doi:10.5291/ILL-DATA.5-31-2717}). 
The excitation spectrum was probed on two time-of-flight spectrometers at the ILL.  Measurements at $T=1.5$ and 100 K with an orange cryostat were performed on PANTHER with incoming energies $E_{\rm i} =$ 19, 35, 60 and 76 meV, the energy resolution at the elastic line being of the order of 4-6\% of $E_i$ \cite{Fak2022} (\href{https://dx.doi.org/10.5291/ILL-DATA.TEST-3151}{doi:10.5291/ILL-DATA.TEST-3151}). To access the low energy part of the spectrum with a high resolution, measurements were performed on SHARP (LLB CRG@ILL) equipped with a dilution fridge with wavelengths of 4.1 and 5.1\ \AA, the latter giving an energy resolution at the elastic line of 70 $\si{\micro\electronvolt}$ (\href{https://dx.doi.org/10.5291/ILL-DATA.CRG-2725}{doi:10.5291/ILL-DATA.CRG-2725}).

\section{\tbscnb\ crystal structure}
\label{AppendixB}

The atomic structure of \tbscnb\ has been refined from a joined analysis of synchrotron based X-ray and neutron diffraction measurements (Table \ref{TableStruct}).

\begin{table}[h!]
 \caption{Results of the joined structural refinement of \tbscnb\ from X-ray (ID22) and neutron (D2B) diffraction data in the Fd$\bar 3$m (227) space group at 300~K with $a$=10.39350(3) \AA\ and a Bragg RF-factor= 8.5 and 4.8 for X-ray and neutron respectively.}
 \label{tab:Ireps}
 \begin{ruledtabular}
 \begin{tabular}{cccccc}
 Atom &Wyckoff & $x$ & $y$ & $z$ & Occ. \\
  \hline
Tb & 16$d$ & 0.5 & 0.5 & 0.5 & 1 \\
 Nb & 16$c$ & 0 & $0$ & 0 & 0.57 \\
 Sc & 16$c$ & 0 & 0 & 0 & 0.43 \\
 O(1) & 48$f$ & 0.3370(1) & 0.1250 & 0.1250 & 1\\
 O(2) & 8$b$ & 0.3750 & 0.3750 & 0.3750 & 1\\
 O(3) & 8$a$ & 0.1250 & 0.1250 & 0.1250 & 0.17\\
  \end{tabular}
 \label{TableStruct}
 \end{ruledtabular}
\end{table}  

\section{Inelastic neutron scattering}
\label{AppendixC}

Additional information is presented concerning the analysis of the inelastic neutron scattering measurements: identification of the magnetic nature of non dispersive excitations (Fig. \ref{fig:INS}) and unfruitful analysis of the low energy excitation with different models (Fig. \ref{fig:INSb}).

\begin{figure}[!h]
\includegraphics[width=8cm]{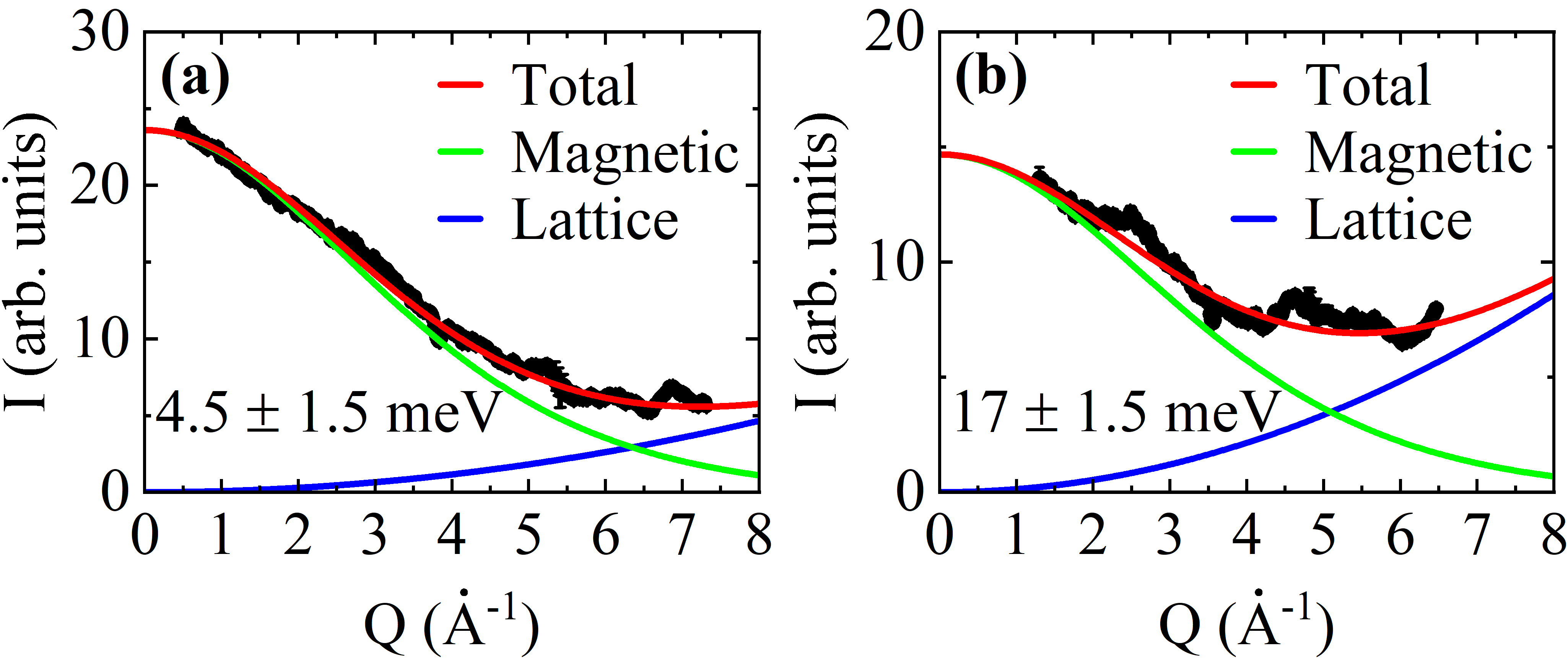}
\caption{Inelastic neutron scattering recorded on PANTHER: $Q$ dependence of the non-dispersive signals at 4.5 (a) and 17~meV (b) integrated on a 3 meV bandwidth around the energy of the excitation. These were fit (red lines) by a sum of two contributions, the squared magnetic form factor of the Tb$^{3+}$ (green line) and the $Q^2$ dependence for the lattice contribution (blue line). It demonstrates that they can be attributed to CEF levels, mixed with phonons at high $Q$.}
\label{fig:INS}
\end{figure}

\begin{figure}[!h]
\includegraphics[width=8cm]{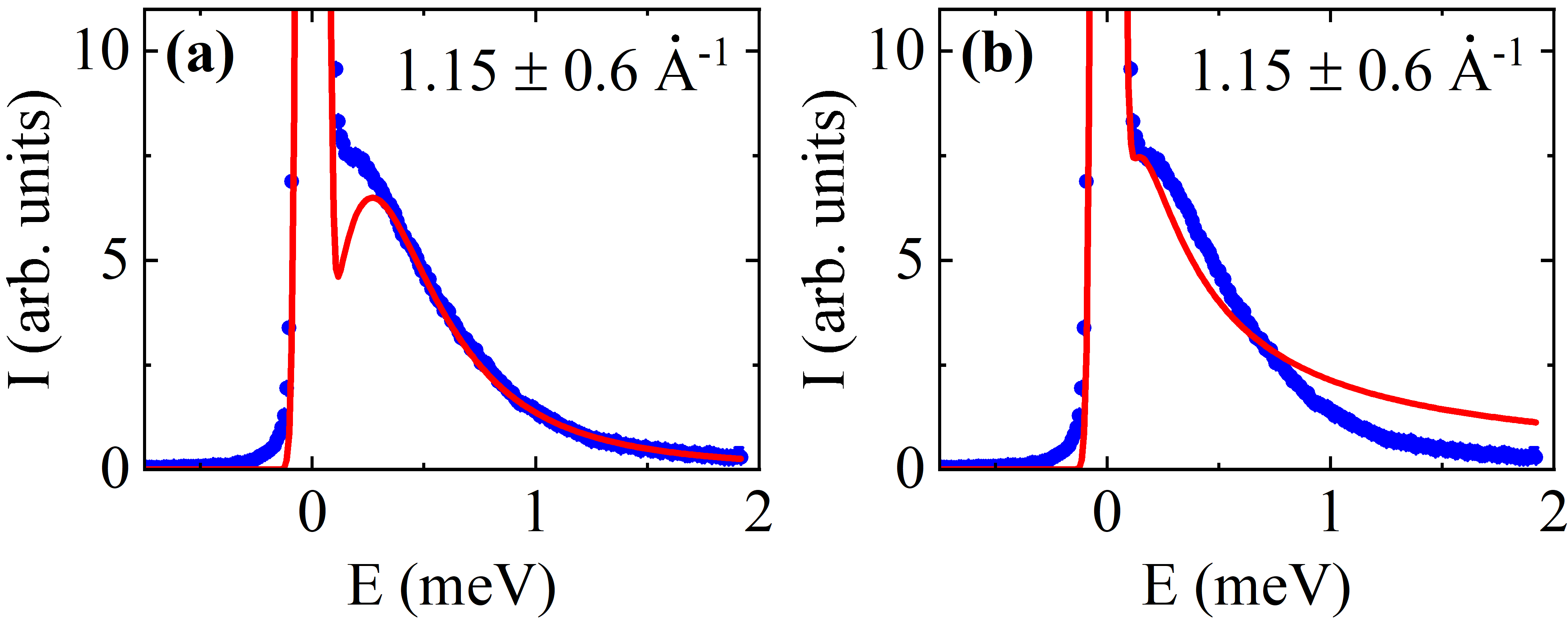}
\caption{Example of different models used to fit the low energy signal measured at 0.05 K by inelastic neutron scattering on SHARP (Fig. \ref{fig:PANTHER}e): (a) A difference of two inelastic Lorentz functions centered at $\pm E_{\mathrm{inel}}= 0.01$ meV and with a full width at half maximum (FWHM) of 0.93 meV to describe a single excitation at finite energy. (b) A quasi-elastic lorentzian function with a 0.29 meV FWHM to describe a quasielastic process. 
Both were multiplied by the detailed balance factor and the elastic peak was accounted for by an additional Gauss function centered at zero energy and of $\mathrm{FWHM}=0.07$~meV corresponding to the experimental resolution. None of these models can reproduce the measurements.}
\label{fig:INSb}
\end{figure}

\section{Point charge model in \tbscnb} 
\label{AppendixD}

For the Tb compound, the crystal field parameters have been obtained from a point charge model for the 13 charge configurations produced by the 57\% Nb / 43\% Sc disorder (Fig. \ref{fig:struct2}d). The percentage of the different $B$ site hexagon configurations were calculated using a Monte Carlo method and a mapping to a spin model on a pyrochlore lattice with a box of $8\times8\times8$ times 16 spins. Assuming a spin ice rule, the starting spin arrangement was calculated with an equal ratio of in and out spins yielding a 2 in 2 out configuration on each tetrahedron. Then a fraction of the spins was flipped in order to obtain a 57\%/43\% ratio. Loop motions and single-spin flips that do not change the proportion of in and out spins were realized, allowing to delocalize randomly the defects. A statistical average of the proportion of the different hexagon configurations was finally obtained using 1000 samplings. The resulting percentages are reported in table \ref{tab:cef_Tb}.

\begin{table}[h!]
\begin{ruledtabular}
\begin{tabular} {cccccccccc}
set & \% & $E_{0}$ & $E_{1}$ & $E_{2}$ & $E_{3}$ & $E_{4}$ & $E_{5}$ & $E_{6}$ & $E_{7}$ \\
\hline\noalign{\vskip 1mm} 
1   &   0.9  & 0 & 0 & 1.472 & 1.472 & 14.115 & 17.125 & 36.117 & 38.067 \\
2   &   0.05  & 0 & 0 & 7.564 & 7.564 & 14.100 & 15.098 & 37.842 & 41.976 \\
3   &   10.1  & 0 & 0.165 & 2.641 & 2.714 & 15.131 & 17.741 & 36.014 & 38.356 \\
4   &   1.4  & 0 & 0.063 & 6.369 & 6.611 & 14.992 & 16.204 & 37.058 & 40.908 \\
5   &   6.34  & 0 & 0.046 & 5.416 & 5.6039 & 15.360 & 16.864 & 36.374 & 39.915 \\
6   & 19.2 & 0 & 0.046 & 5.416 & 5.6039 & 15.360 & 16.864 & 36.374 & 39.915 \\
7   &   9.5 & 0 & 0.182 & 5.178 & 5.667 & 16.19 & 17.625 & 36.741 & 40.197 \\
8   &   2.4 & 0 & 0.132 & 3.446 & 3.634 & 15.437 & 17.660 & 35.828 & 38.576 \\
9   &   7.2 & 0 & 0.132 & 3.446 & 3.634 & 15.437 & 17.660 & 35.828 & 38.576 \\
10 &   3.5 & 0 & 0.161 & 3.837 & 4.079 & 16.356 & 18.507 & 36.362 & 39.135 \\
11 &   3.9  & 0 & 0 & 4.457 & 4.457 & 15.201 & 17.079 & 35.792 & 38.962 \\
12 & 23.5 & 0 & 0.146 & 4.431 & 4.769 & 16.094 & 17.889 & 36.247 & 39.390 \\
13 & 12.0 & 0 & 0 & 4.457 & 4.457 & 15.201 & 17.079 & 35.792 & 38.962 \\
\end{tabular}
\end{ruledtabular}
\caption{Percentage of occurence and energies (in meV) of the crystal field levels in \tbscnb\ for the 13 configurations calculated with a unique set of shielding factors and 57 \% Nb / 43\% Sc.}
\label{tab:cef_Tb}
\end{table}

The crystal field Hamiltonian is given below:
\begin{equation} 
\begin{aligned}
\widehat{\mathcal{H}}_{\mathrm{cf}}= \sum_{n,m} B_{n}^{m}\widehat{\mathcal{O}}_{n}^{m}.
\end{aligned}
\label{eq:Hcf}
\end{equation}
with $n$=2, 4, 6 and $m\leq6$. $\widehat{\mathcal{O}}_{n}^{m}$ are the Wybourne operators and $B_{n}^{m}$ are crystal field parameters, which are determined in first approximation by the charge environment (positions and charges of the neighbors) of the rare earth Tb$^{3+}$ ($L=3$, $S=3$, $g_\mathrm{J}J=9$).

For the non-distorted $\mathcal{D}_\mathrm{3d}$ Tb$^{3+}$  site symmetry, only $B_{2}^{0}$, $B_{4}^{0}$, $B_{4}^{3}$, $B_{6}^{0}$, $B_{6}^{3}$, $B_{6}^{6}$ terms are non zero, while other terms are relevant for lower symmetry. Additional shielding factors are usually necessary to obtain the correct CEF scheme \cite{Sternheimer1968}. We adjusted these shielding factors for each type of neighbors in order to reproduce the first CEF levels at $\approx$ 4.5 and $\approx$ 17 meV for a fictitious environment with an average charge of +4 on the $B$ sites. The shielding factors (87.1~\%, 88.35\%, 97.5 \% and  and 97.9 \% for 8b O$^{2-}$, 48f O$^{2-}$, 16d Nb$^{5+}$/Sc$^{3+}$ and 16c Tb$^{3+}$ respectively) were then used to calculate the crystal field parameters and CEF energy scheme for each of the 13 configurations. The latter are reported in Table \ref{tab:cef_Tb}. We further calculated the single-ion magnetic susceptibility (Fig. \ref{fig:invchi}) and specific heat [Fig. \ref{fig:Cpchi}b] using these CEF schemes and the percentage of each configuration given in Table \ref{tab:cef_Tb}.
 
\begin{figure}[!h]
\centering
\includegraphics[width=\columnwidth]{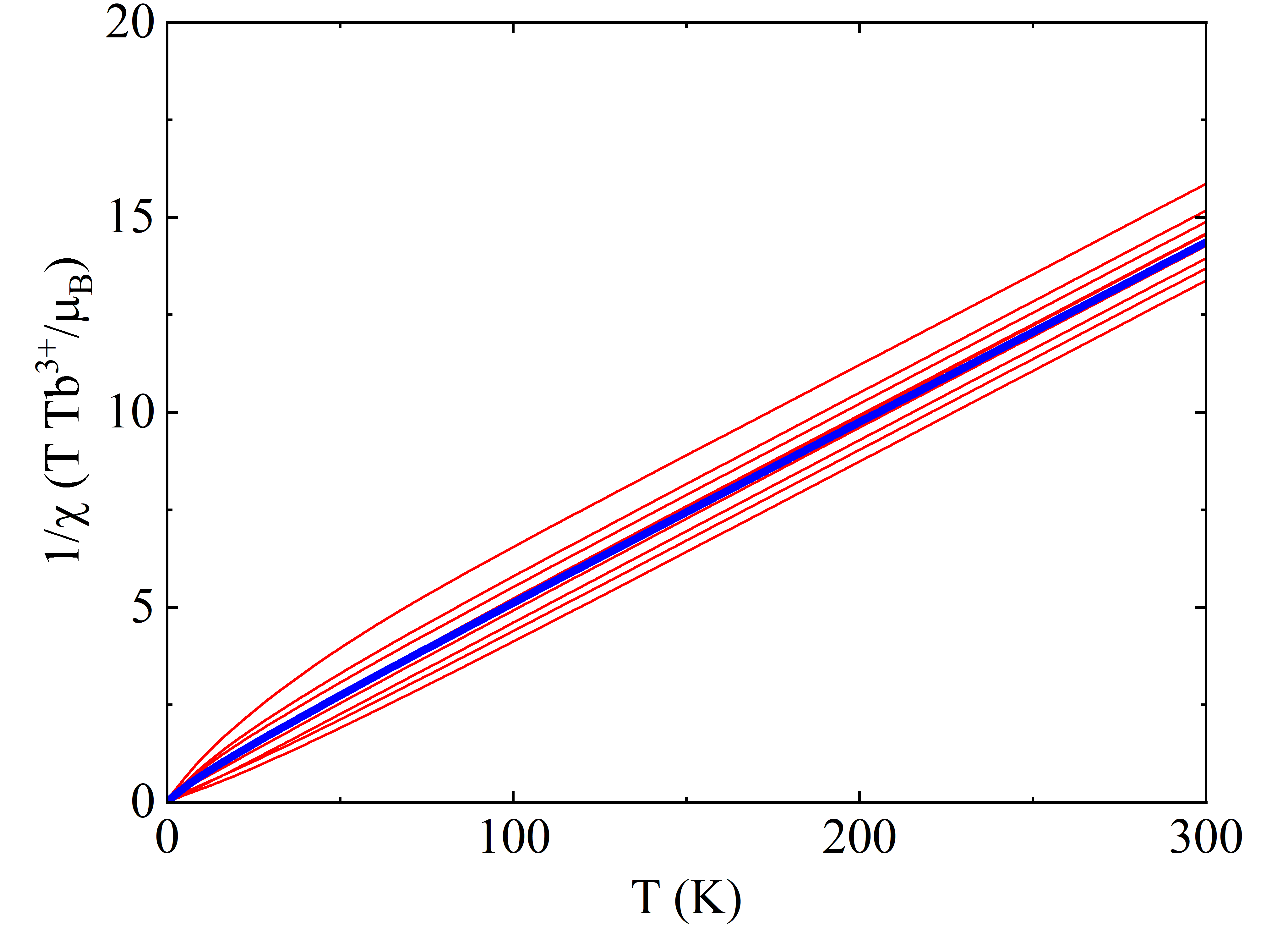}
\caption{Calculated inverse magnetic susceptibility of \tbscnb\ from the 13 configurations of the disordered point charge model (red lines) and the resulting weighted average assuming 57\% Nb$^{5+}$ and 43\% Sc$^{3+}$ (blue line). This shows the influence of the charge disorder on the single-ion magnetic susceptibility: the curves exhibit a range of intercepts with the temperature axis between $-40$ and $+15$~K despite the absence of interactions in the model, thus making a Curie-Weiss analysis of the experimental data unsuitable to extract the strength of magnetic interactions.}
\label{fig:invchi}
\end{figure}

\section{Macroscopic measurements in \tbscnb}
\label{AppendixE}

A small hysteresis cycle in the isothermal magnetization curves at low temperature is shown in Fig. \ref{fig:hysteresis}. The failure of the analysis of the AC susceptibility measurements using an Arrhenius law is discussed in the caption of Fig \ref{fig:Arrhenius}. 

\begin{figure}[h!]
\includegraphics[width=\columnwidth]{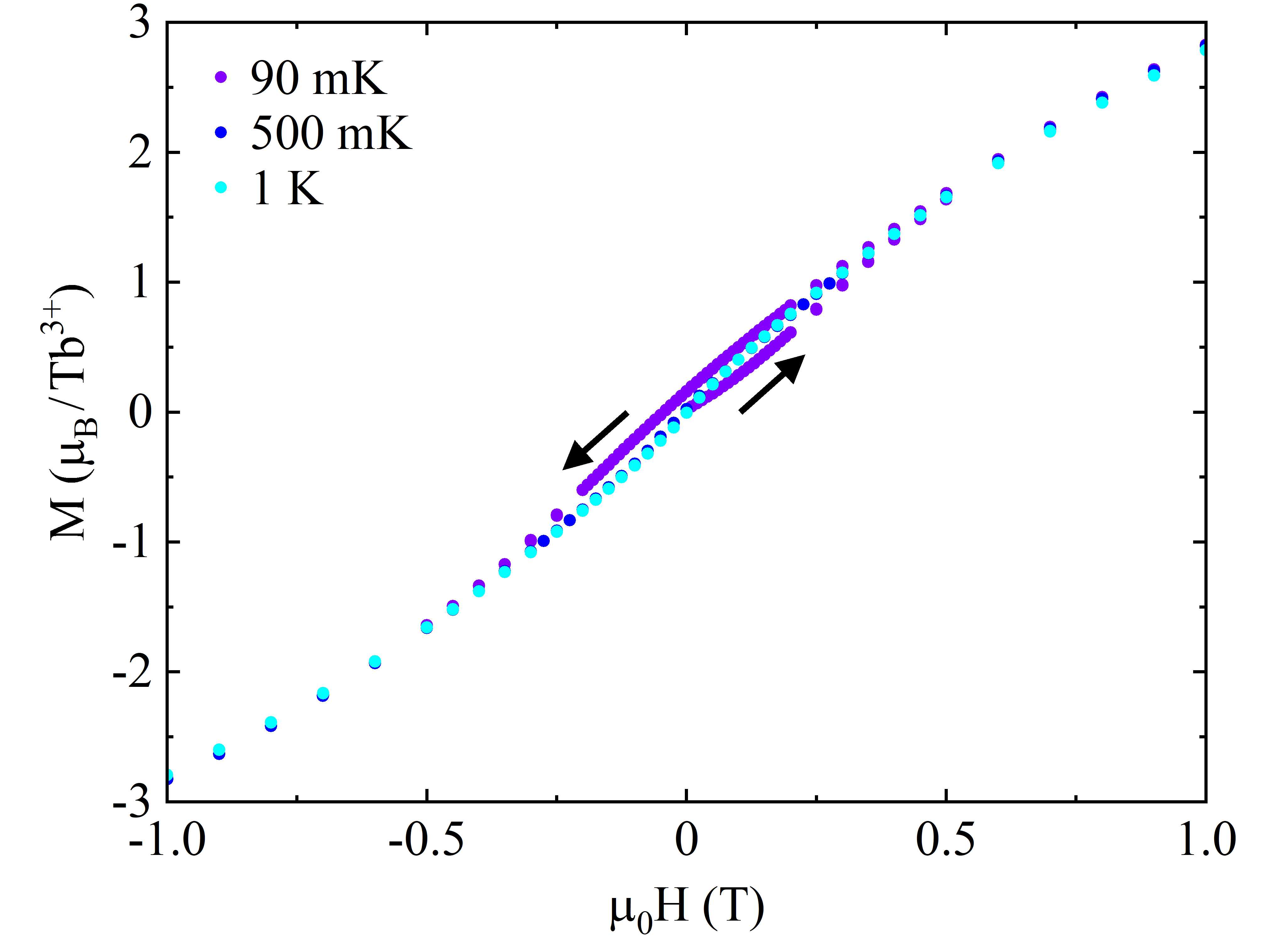}
\caption{\tbscnb\ magnetization as a function of increasing and decreasing magnetic fields at different temperatures between 0.09 and 1 K showing an hysteresis cycle at low temperature.}
\label{fig:hysteresis}
\end{figure}

\begin{figure}[h!]
\includegraphics[width=\columnwidth]{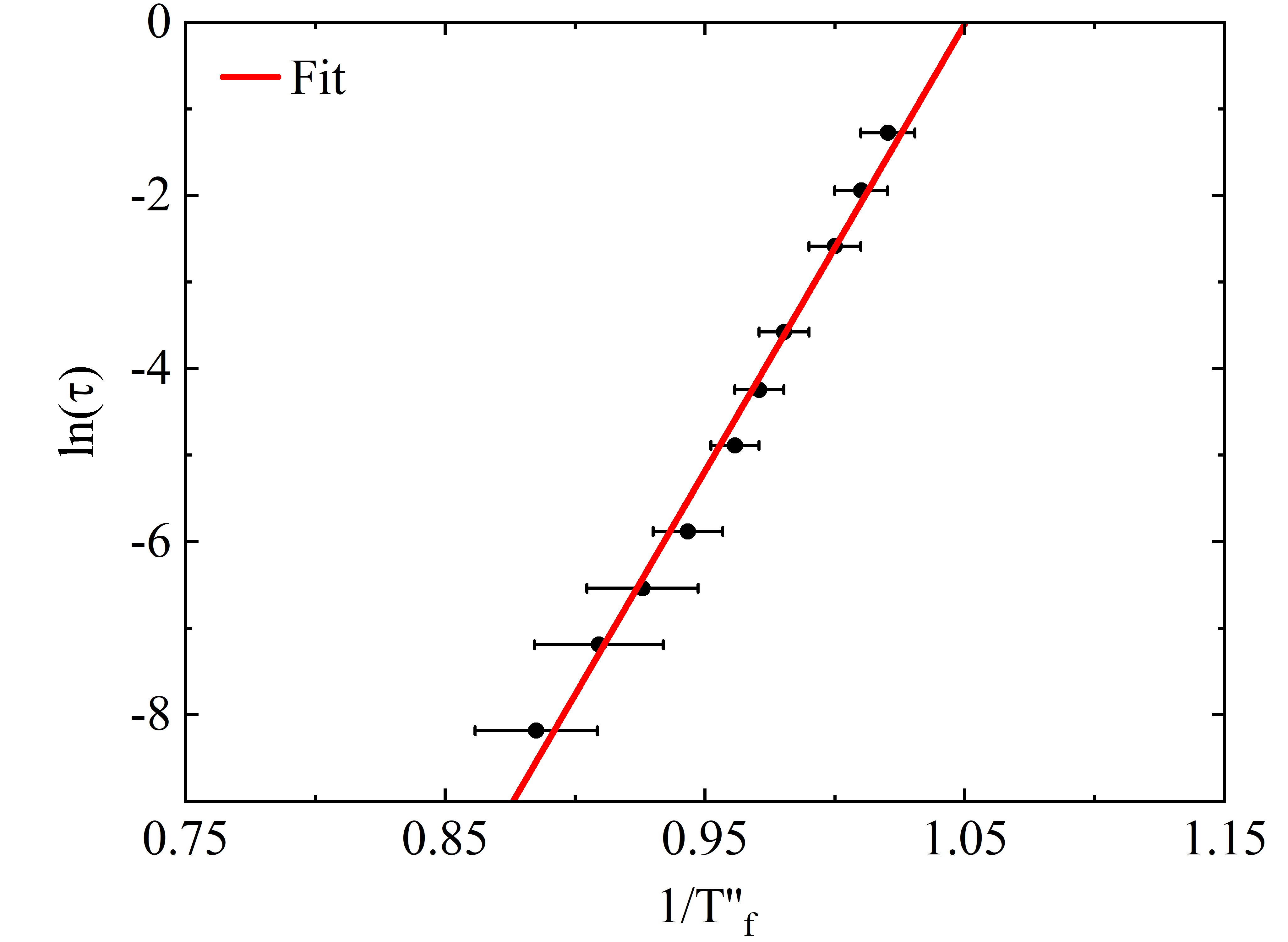}
\caption{$\ln\tau$ versus $1/T''_f$ from AC susceptibility measurements (black dots). The relaxation time $\tau=1/2\pi f$ was obtained from the maximum of $\chi''$ vs $T$ measurements at a fixed frequency $f$ (Fig. \ref{fig:ZFCchiAC}c). The red line in (b) is a fit to an Arrhenius law $\tau=\tau_0 \exp(E_a/T)$ yielding $\tau_0=3.5 \times 10^{-24}$ s and the energy barrier $E_a= 52$ K. The obtained value for the shortest relaxation time available to the system $\tau_0$ is unphysically too low. This rules out a thermally activated process as the origin of the observed irreversibilities.}
\label{fig:Arrhenius}
\end{figure}

\section{\prscnb}
\label{AppendixF}

A polycrystalline sample of \prscnb\ (Appendix \ref{AppendixA}) was measured by neutron diffraction on D1B using a dilution insert allowing to reach 50 mK with a wavelength of 2.52 \AA\ (\href{https://dx.doi.org/10.5291/ILL-DATA.EASY-860}{doi:10.5291/ILL-DATA.EASY-860}). No difference is observed between 50 K and 50~mK. The Rietveld refinement of the D1B data at 50 K yields a lattice parameter $a$=10.5218(2) \AA, a 48f oxygen coordinate $x$=0.3295(2) and an occupancy of the $B$ site by 59\% of Nb$^{5+}$ and 41\% of Sc$^{3+}$ (Fig. \ref{fig:PrStruc}). 

Macroscopic measurements are shown in Fig. \ref{fig:Pr}. The inverse susceptibility (panel a) is not linear, even at ``high" temperature, which suggests the presence of crystal electric field levels in the 10-20 meV energy range. This is consistent with previous reports on \prsn\ \cite{Princep2013}, \przr\ \cite{Kimura2013} and \prhf\ \cite{Sibille2016, Anand2016}. 

\begin{figure}[h!]
\includegraphics[width=\columnwidth]{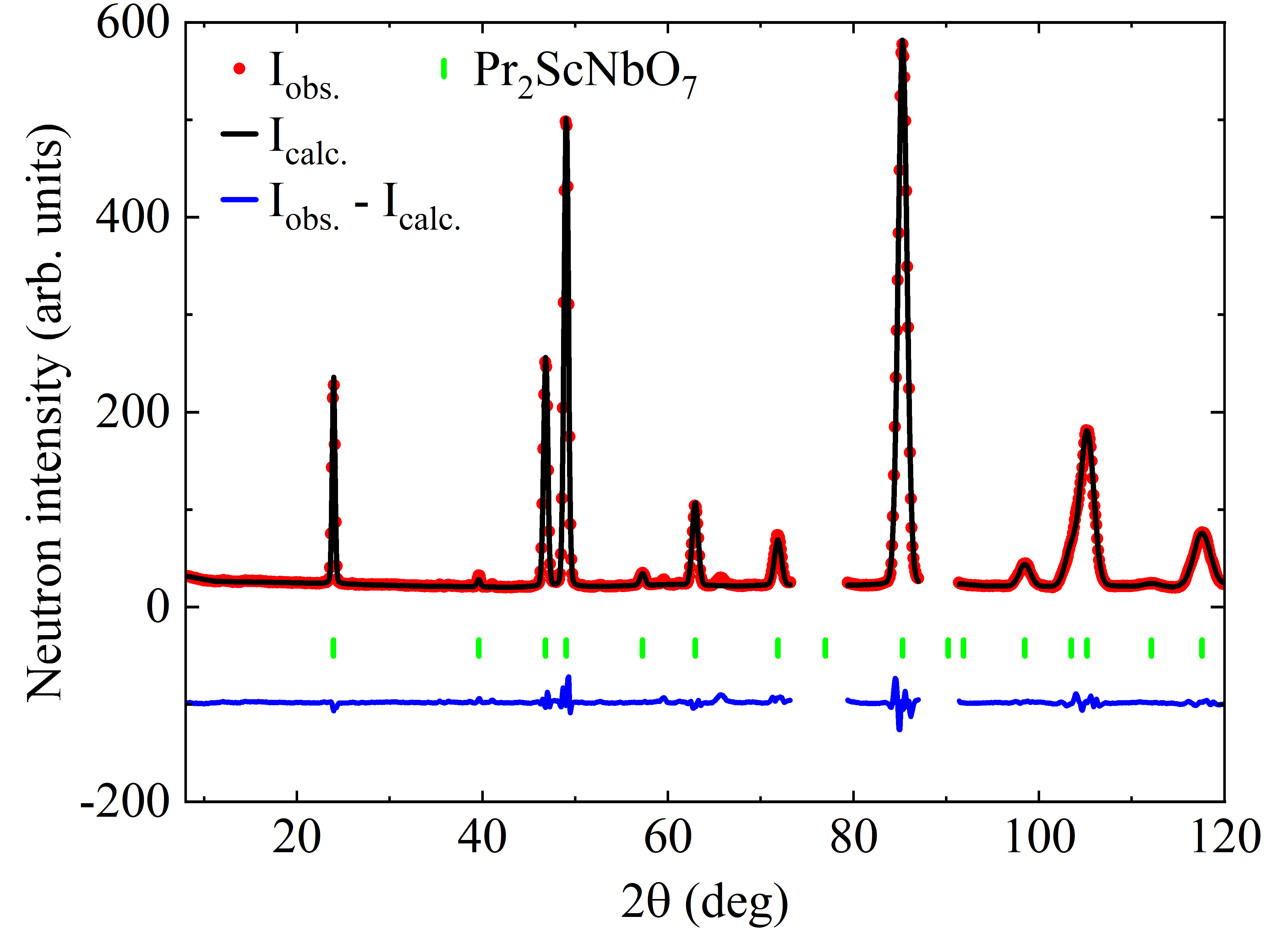}
\caption{Neutron diffractogram of the \prscnb\ polycrystalline sample measured on D1B at 50 K. The measurement is in red, the Rietveld refinement in black and the difference in blue. The Bragg $R$-factor is 3.}
\label{fig:PrStruc}
\end{figure}

\begin{figure}
\includegraphics[width=9cm]{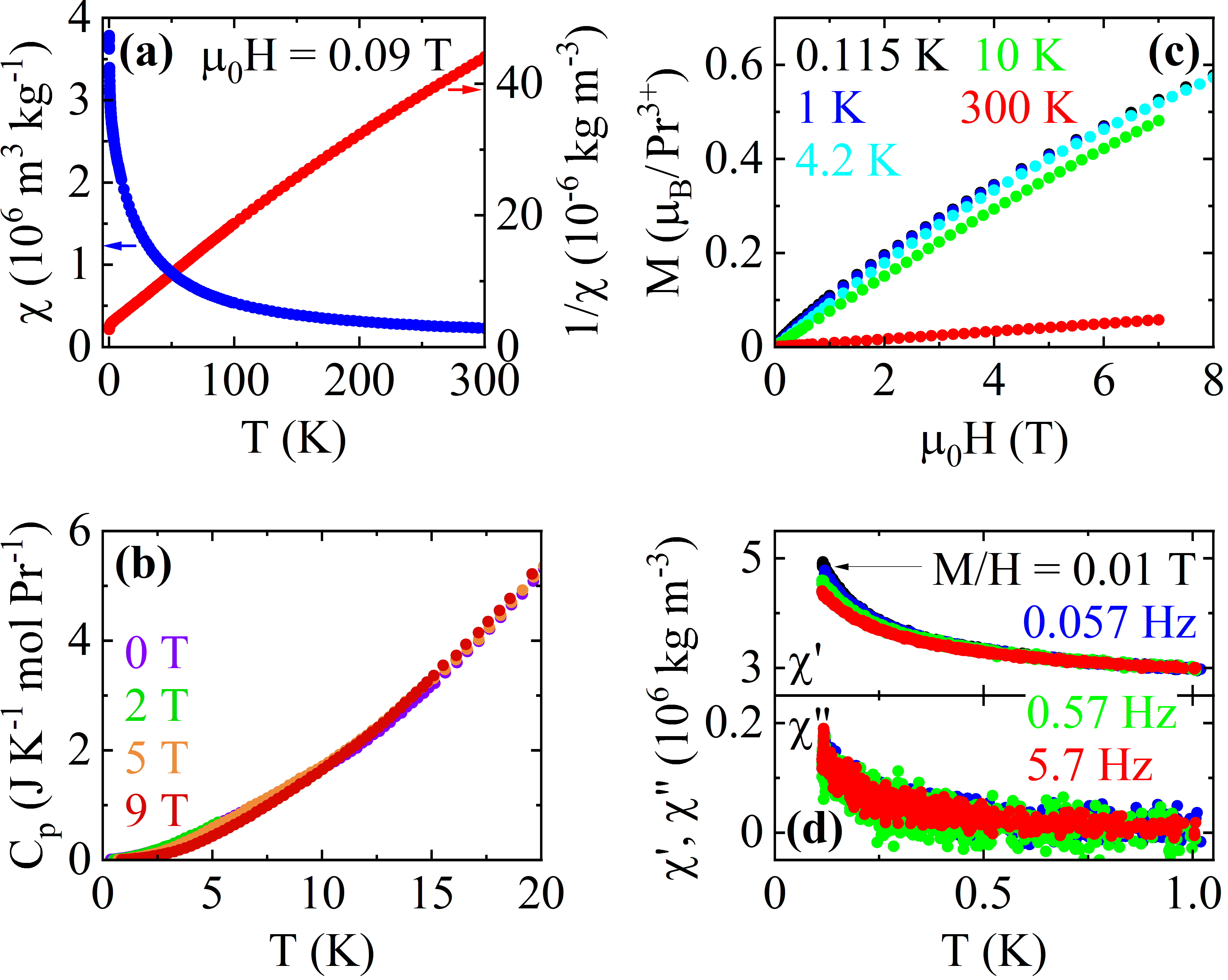}
\caption{Macroscopic measurements on a \prscnb\ polycrystalline sample. (a) Inverse magnetic susceptibility versus temperature measured in a magnetic field of 0.09 T. (b) Specific heat versus temperature for different magnetic fields. (c) Magnetization curves versus magnetic field at various temperatures. (d) Real and imaginary parts of the AC susceptibility versus temperature at three different frequencies. }
\label{fig:Pr}
\end{figure}

The specific heat at low temperature is however different from these compounds. While a bump was observed in other Pr based systems, a monotonous behavior is observed in \prscnb\ from 400 mK to 20 K (Fig. \ref{fig:Pr}b). This is in agreement with both the absence of significant magnetic interactions in this energy range as inferred from neutron scattering and the absence of low energy CEF crystal field levels. 

We modeled the splitting of the ground state doublet with a point charge model assuming 59\% Nb$^{5+}$ and 41\% Sc$^{3+}$ combined to a charge ice rule and adjusting the charge shielding factors such as to reproduce the CEF of the \prhf\ compound \cite{Sibille2016, Anand2016} (Table \ref{tab:cef_Pr}). The shielding factors are 93\%, 83\%, 95 \% and  and 96.1 \% for 8b O$^{2-}$, 48f O$^{2-}$, 16d Nb$^{5+}$/Sc$^{3+}$ and 16c Pr$^{3+}$ respectively. This model produces a splitting of the ground state of the order of 3 $-$ 4 meV in average, which is expected to yield a peak in specific heat at temperatures higher than 20 K. 

 \begin{table}[h!]
\begin{ruledtabular}
\begin{tabular} {cccccccccc}
set &  $E_{0}$ & $E_{1}$ & $E_{2}$ & $E_{3}$ & $E_{4}$ & $E_{5}$ & $E_{6}$ & $E_{7}$  \\
\hline\noalign{\vskip 1mm} 
1     & 0 & 0 & 36.62 & 60.07 & 60.07 & 78.20 & 99.40 & 123.62 \\
2     & 0 & 0 & 14.05 & 55.87 & 55.87 & 61.80 & 99.85 & 104.93\\
3     & 0 & 5.12 & 10.38 & 47.44 & 52.72 & 62.15 & 94.93 & 102.78 \\
4     & 0 & 2.82 & 27.98 & 56.03 & 68.19 & 77.11 & 98.73 & 118.26 \\
5    & 0 & 2.98 & 18.94 & 54.28 & 57.71 & 72.14 & 95.33 & 112.86 \\
6     & 0 & 2.98 & 18.94 & 54.28 & 57.71 & 72.14 & 95.33 & 112.86 \\
7      & 0 & 7.07 & 19.50 & 47.70 & 66.96 & 75.92 & 101.39 & 114.64 \\
8     & 0 & 3.71 & 5.12 & 44.86 & 55.45 & 58.10 & 91.89 & 102.02 \\
9     & 0 & 3.71 & 5.12 & 44.86 & 55.45 & 58.10 & 91.89 & 102.02 \\
10    & 0 & 2.29 & 9.04 & 42.83 & 49.71 & 72.53 & 94.93 & 106.70 \\
11     & 0 & 0 & 9.16 & 9.16 & 51.58 & 51.58 & 90.37 & 104.87 \\
12   & 0 & 6.47 & 11.39 & 45.95 & 56.82 & 72.13 & 96.37 & 109.00\\
13   & 0 & 0 & 9.16 & 51.58 & 51.58 & 55.72 & 90.37 & 104.87\\
\end{tabular}
\end{ruledtabular}
\caption{Energies (in meV) of the crystal field levels in Pr$_2$NbScO$_7$ for the 13 configurations calculated with a unique set of shielding factors.}
\label{tab:cef_Pr}
\end{table}

The isothermal magnetization curves (Fig. \ref{fig:Pr}c) are very smooth and the magnetization in an 8 T applied field does not evolve below 4.2 K, reaching the low value of 0.6 $\mu_{\rm B}/$Pr. Only AC susceptibility yields possible signatures of weak magnetic interactions: a slight onset of $\chi''$ is observed below 200 mK, concomitantly with signatures of a frequency dependence in $\chi'$ (Fig. \ref{fig:Pr}d). These observations point to weak magnetic interactions between the diluted magnetic moments, possibly from dipolar origin, and undetected through neutron scattering. These results are very similar to those reported in Pr$_2$InSbO$_7$ \cite{Ortiz2022} where a trivial non-magnetic singlet state was proposed. 

In addition to the  \prscnb\ polycrystalline sample, we also synthesized a single crystal by the floating zone method under flowing air in a Cyberstar furnace. X-ray diffraction and transmission electron microscopy were performed on crushed single crystal parts to ensure the good crystallinity of the pyrochlore structure and the absence of superstructure. The good quality of the single crystal was also assessed by Laue diffraction. The color of the single crystal is red, at variance with the green color of other Pr-based pyrochlores \cite{CiomagaHatnean2017, Koohpayeh2014}. We did not investigated further this single crystal in view of the results on the powder sample.


%

\end{document}